\documentclass[fleqn,10pt]{wlscirep}
\usepackage[utf8]{inputenc}
\usepackage[T1]{fontenc}

\usepackage{amssymb,amsmath,multirow,rotate,color}
\usepackage{setspace}
\usepackage{BOONDOX-uprscr}
\usepackage[mathlines]{lineno} 

\usepackage{url}

\renewcommand{\url}[1]{\href{https://doi.org/#1}{#1}}

\usepackage[defaultcolor=black]{changes}

\title{\added{A unified framework for n}et effects in signed social and ecological networks}

\author[1,2]{Carlos Gómez-Ambrosi}
\author[3,4,*]{Violeta Calleja-Solanas}
\affil[1]{Department of Mathematics, University of Zaragoza, 50009 Zaragoza, Spain}
\affil[2]{GOTHAM lab, Institute for Biocomputation and Physics of Complex Systems (BIFI), University of Zaragoza, 50018 Zaragoza, Spain}
\affil[3]{Department of Biology, University of Oxford, Life and Mind Building, Oxford OX1 3PS, United Kingdom}
\affil[4]{Department of Ecology and Evolution, Doñana Biological Station (EBD-CSIC), 41092 Seville, Spain}
\affil[*]{violeta.calleja@biology.ox.ac.uk}

\keywords{signed networks, net effects, indirect effects, Lotka-Volterra dynamics, Katz centrality, PageRank}

\begin{abstract}
With improvements in data resolution and quality, researchers can now represent complex systems as signed, weighted, and directed networks. In this article, we introduce a framework for measuring net and indirect effects without simplifying these information-rich networks. It captures both direct and indirect interactions, the effect of the whole network on a node, and conversely, the effect of a node on the entire network, while accommodating the complexity of signed, weighted, and directed links. Our taxonomy unifies and extends existing approaches and measures from network science, computational social science, and ecological networks. We demonstrate its value in ecological systems, where net and indirect effects are critical yet difficult to quantify. Using generalized Lotka–Volterra dynamics, we find a strong correlation between negative net effects and species extinction. We further apply the framework to a real-world social network, where it identifies informative rankings that illuminate influence propagation and power dynamics.
\end{abstract}

\begin{document}

\flushbottom
\maketitle
%
%
\thispagestyle{empty}

\section*{Introduction}

Complex networks provide a powerful framework for encoding interactions between entities, represented as nodes and links.
\added{In ecological networks, for instance, species serve as nodes, while links can represent co-occurrence, carbon flow, or visitation frequency. When studied under Lotka–Volterra models, links have the potential to quantify the strength and sign of interactions. Specifically, a directed link from species $j$ to species $i$ is weighted by the interaction coefficient, $a_{ij}$, which quantifies the direct positive or negative effect of species $j$ on the per capita growth rate of species $i$~\cite{novak2016}.}
Similarly, in social networks, nodes often represent agents, while signed weighted links characterize the strength and nature of their relationships. The structure of these networks is important to characterize because it affects their functioning. In ecology, interaction networks underpin fundamental processes such as species coexistence~\cite{chesson2000} and biodiversity maintenance~\cite{godoy2018}. In social systems, network topology affects contagion processes~\cite{soriano2018} and the dynamics of influence and power \cite{vendeville2025}, among other topics. 

A common strategy to characterize network structure involves computing a centrality measure that assesses the importance of nodes within the network. There are many possible definitions of importance in the literature. Examples include closeness centrality, which captures how efficiently a node can access others, and betweenness centrality, which measures the extent to which a node acts as a bridge between other nodes~\cite{newman2018}. Other centralities, such as Katz's~\cite{katz1953} and PageRank \cite{gleich2015}, incorporate the influence of network structure to assess the importance of nodes. However, their application is often limited to unsigned and unweighted networks ---constraining their relevance for many real-world systems---  because of the mathematical complexity of dealing with signed, weighted, and directed networks simultaneously. Signed networks, where links can represent positive or negative interactions (e.g., predation in ecological networks or trust in social networks), require a more nuanced interpretation. Weighted networks, which assign varying strengths to interactions, add an additional layer of complexity. Similarly, the directionality of links, essential for understanding cause-and-effect relationships, is typically underexplored in combination with sign and weight~\cite{newman2018}. While existing studies \cite{everett2014,liu2020} have attempted to extend these measures to incorporate such features, the proposed indices were tailored specifically for their study systems. Significant gaps remain, particularly in their ability to provide a complete picture of node influence applicable to different systems and research questions.

An equally important limitation of current network measures lies in their difficulty in capturing the total net influence, which aggregates both direct and indirect effects.
In many complex networks, the influence of a node extends beyond its immediate connections, propagating through paths of varying length and significance. For instance, ignoring indirect ecological interactions can lead to oversights in conservation strategies or misinterpretations of ecosystem stability~\cite{wootton1994,guimaraes2017}. In social systems, indirect effects are central to phenomena like the spread of influence, where a person’s actions may have a ripple effect on their extended network~\cite{omodei2015}. Current centrality measures, such as Katz's and PageRank, account for indirect effects only in the context of the cumulative effect of the network on a single node and cannot disentangle their contribution to the final measure. By doing so, we risk overlooking the full extent of node influence. 

To address these limitations, we propose shifting the focus from traditional centrality to the more comprehensive notion of net effect. Intuitively, net effects are the overall, aggregated effects through both direct and indirect pathways. Even though this verbal definition may seem simple, it raises follow-up questions ---the net effect of one node on another, the influence a particular node exerts over the entire network, or the collective impact of all nodes on a single target--- that do not depend on a particular definition of importance. This perspective allows us to capture the interconnected nature of networks and its consequences on their dynamics. 

Here, we develop a framework to compute a matrix of net effects for signed, weighted, and directed networks. We present a taxonomy of measures along with their mathematical details that enable us to (i) measure the effect of the network on a node, (ii) quantify the reverse measure, i.e.~the effect of a node on the network, and (iii) disentangle direct effects from indirect effects (Fig.~\ref{fig:1}).
We also generalize centrality measures ---such as Katz's and PageRank--- to signed, weighted, and directed networks. These tools can be applied to a broad range of networks. In particular, we contextualize net effects for social and ecological networks, clarifying the correspondences and relations among other isolated measures of net effects, and providing some direct applications. We aim to unify terminology and make clear the possible characterizations of net effects. 

\section*{Results}

\subsection*{Measures of net effects in signed weighted directed networks} \label{measures}

\begin{figure*}
    \centering
    \includegraphics[width=\linewidth]{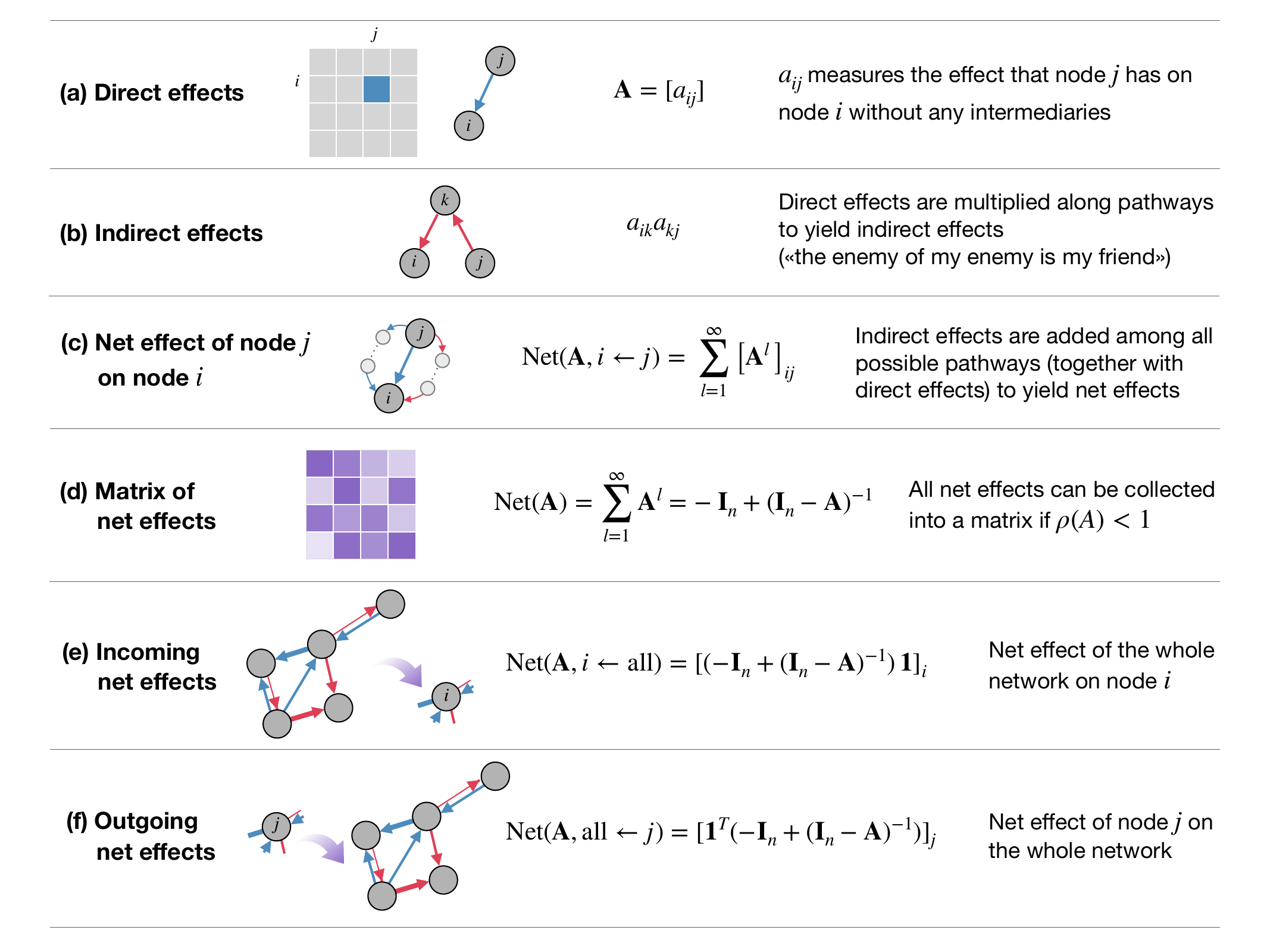}
    \caption{Overview of the fundamental notions developed for net effects. Arrow color and thickness represent the sign and weight of the links.}
    \label{fig:1}
\end{figure*}

Our starting point is a signed weighted directed network with $n$ nodes, represented by its $n \times n$ (signed weighted) adjacency matrix $\mathbf{A} = [a_{ij}]$.
Entry $a_{ij}$ of $\mathbf{A}$ can take any value, and represents the \textbf{direct effect} of node $j$ on node $i$ (Fig.~\ref{fig:1}a).
Our intended interpretation is that the nodes in the network interact, and that $a_{ij}$ measures the strength of the interaction $i \leftarrow j$ (including whether this interaction is positive or negative).
Since $\mathbf{A}$ represents direct effects or interaction strengths rather than mere adjacency, we call it the matrix of direct effects or the interaction matrix of the network.

We will say that a network is simple (as in~\cite{newman2018}, p.106) if it does not contain self-loops. 
Equivalently, its interaction matrix has zero diagonal, i.e.~$a_{ii} = 0$ for all $i$.
\added{While certain interesting results are valid specifically for simple networks, as discussed below, we do not assume our networks to be simple in general.}

If $i \leftarrow k \leftarrow j$ is a walk of length $2$ from node $j$ to node $i$, with node $k$ as its middle node, then the \textbf{indirect effect} of node $j$ on node $i$ along this walk is defined as $a_{ik} a_{kj}$  (Fig.~\ref{fig:1}b), i.e.~direct effects are multiplicative along walks.
This complies with the rule that "the enemy of my enemy is my friend", in accordance with the usual interpretation in social and ecological networks (for the latter, see e.g.~\cite{case2000}, p.363).
In general, if $i \leftarrow k_{\ell-1} \leftarrow \cdots \leftarrow k_2 \leftarrow k_1 \leftarrow j$ is a walk of length $\ell > 1$ from node $j$ to node $i$, then the indirect effect of node $j$ on node $i$ along this walk is defined as $a_{ik_{\ell-1}} \cdots a_{k_2k_1} a_{k_1j}$.

Continuing in this way, we define the indirect effect of node $j$ on node $i$ of order $\ell > 1$ as the sum of the indirect effects of node $j$ on node $i$ along all walks of length $\ell$ from node $j$ to node $i$, so that indirect effects are additive among walks.
Notice that this is precisely the $(i,j)$-entry of the $\ell$-th power of matrix $\mathbf{A}$:
\begin{equation}
[\mathbf{A}^\ell]_{ij} =
\sum_{k_1,\ldots,k_{\ell-1}=1}^n a_{ik_{\ell-1}} \cdots a_{k_2k_1} a_{k_1j}.
\end{equation}
Hence $\mathbf{A}^\ell$ is the matrix of indirect effects of order $\ell > 1$, and direct effects can be considered as indirect effects of order $1$.

Finally, we define the net effect of node $j$ on node $i$ as the sum of the indirect effects of node $j$ on node $i$ of all orders, i.e.
\begin{equation}
\sum_{\ell = 1}^{\infty} \, [\mathbf{A}^\ell]_{ij},
\end{equation}
provided that this series converges.
The convergence of this number series for all $i$ and $j$ is equivalent to the convergence of the matrix series
\begin{equation} \label{matrixseries}
\sum_{\ell = 1}^{\infty} \mathbf{A}^\ell,
\end{equation}
in which case we have the equality
\begin{equation} \label{aside}
\sum_{\ell = 1}^{\infty} \, [\mathbf{A}^\ell]_{ij} =
[ \, \sum_{\ell = 1}^{\infty} \mathbf{A}^\ell \, ]_{ij}.
\end{equation}

A small aside is in order here.
If direct effects are non-dimensional, then indirect effects are non-dimensional as well, and they can be added up to yield a non-dimensional net effect.
But if direct effects have units, say $u$, then indirect effects of order $\ell$ have units $u^\ell$, and adding indirect effects of different orders does not have any <<physical>> meaning.
In that case, interaction strengths can be transformed into relative interaction strengths in meaningful ways, so that they become non-dimensional (as in Section "The alpha matrix").
See~\cite{zelnik2024} for more details on this issue.

Returning to the definition of net effects, to characterize the convergence of the matrix series in Eq.~\eqref{matrixseries}, let $\mathbf{I}_n$ denote the $n \times n$ identity matrix and let $\rho (\boldsymbol{\cdot})$ denote the spectral radius of a matrix.
By a well-known result in matrix analysis (see e.g.~\cite{meyer2000}, p.618), the Neumann series $\mathbf{I}_n + \mathbf{A} + \mathbf{A}^2 + \cdots$ converges if and only if $\rho (\mathbf{A}) < 1$, in which case $(\mathbf{I}_n - \mathbf{A})^{-1}$ exists and has a Neumann series expansion
\begin{equation}
(\mathbf{I}_n - \mathbf{A})^{-1} =
\sum_{\ell = 0}^{\infty} \mathbf{A}^\ell.
\end{equation}
Notice that this expansion adds an extra term $\mathbf{I}_n$ to the matrix series in Eq.~\eqref{matrixseries}.
Mathematically, $\mathbf{I}_n = \mathbf{A}^0$ can be considered as the matrix of indirect effects of order $0$.

Therefore, under the hypothesis that $\rho (\mathbf{A}) < 1$, there is a well-defined \textbf{matrix of net effects} of $\mathbf{A}$ (Fig.~\ref{fig:1}d), given by
\begin{equation} 
\mathrm{Net}(\mathbf{A}) =
- \mathbf{I}_n + (\mathbf{I}_n - \mathbf{A})^{-1} = \sum_{\ell = 1}^{\infty} \mathbf{A}^\ell,
\end{equation}
and a well-defined matrix of <<\textbf{plus one}>> net effects of $\mathbf{A}$, given by
\begin{equation}
\mathrm{Net}_{+1}(\mathbf{A}) =
\mathrm{Net}(\mathbf{A}) + \mathbf{I}_n =
(\mathbf{I}_n - \mathbf{A})^{-1} = \sum_{\ell = 0}^{\infty} \mathbf{A}^\ell.
\end{equation}
The \textbf{net effect} of node $j$ on node $i$ (Fig.~\ref{fig:1}c) is then the $(i,j)$-entry of $\mathrm{Net}(\mathbf{A})$, i.e.
\begin{equation} \label{eq:net}
\mathrm{Net}(\mathbf{A},i \leftarrow j) =
\mathrm{Net}(\mathbf{A})_{ij} =
\sum_{\ell = 1}^{\infty} \, [\mathbf{A}^\ell]_{ij},
\end{equation}
and similarly for the plus one net effect of node $j$ on node $i$.
Net effects as in Eq.\eqref{eq:net} are more natural than plus one net effects, but we need to consider the latter in order to include some of the classical centrality measures in our framework, as we shall see below.

In some situations, it is interesting to consider the joint net effect that all nodes have on node $i$.
Thus we define the \textbf{incoming net effect} on node $i$ (Fig.~\ref{fig:1}e) by the expression
\begin{equation}
\mathrm{Net}(\mathbf{A},i \leftarrow \mathrm{all}) =
\sum_{j = 1}^n \mathrm{Net}(\mathbf{A},i \leftarrow j),
\end{equation}
which measures the net effect of the whole network on node $i$.
If we collect all the incoming net effects on a single (column) vector, we get the vector of row sums of the matrix of net effects:
\begin{equation}
\mathrm{Net}(\mathbf{A},* \leftarrow \mathrm{all}) =
\mathrm{Net}(\mathbf{A}) \, \mathbf{1} =
(- \mathbf{I}_n + (\mathbf{I}_n - \mathbf{A})^{-1}) \, \mathbf{1},
\end{equation}
where $\mathbf{1}$ represents the (column) vector of all ones.

We can similarly define the \textbf{outgoing net effect} of node $j$ (Fig.~\ref{fig:1}f) by the expression
\begin{equation}
\mathrm{Net}(\mathbf{A},\mathrm{all} \leftarrow j) =
\sum_{i = 1}^n \mathrm{Net}(\mathbf{A},i \leftarrow j),
\end{equation}
which measures the net effect of node $j$ on the whole network.
If we collect all the outgoing net effects on a single row vector, we get the vector of column sums of the matrix of net effects:
\begin{equation}
\mathrm{Net}(\mathbf{A},\mathrm{all} \leftarrow *) =
\mathbf{1}^T \mathrm{Net}(\mathbf{A}) =
\mathbf{1}^T (- \mathbf{I}_n + (\mathbf{I}_n - \mathbf{A})^{-1}),
\end{equation}
where a $T$ superscript denotes transposition.

All these notions have their plus one counterparts, which are defined straightforwardly.
The following are some relations between net effects and plus one net effects:
\begin{equation}
\mathrm{Net}_{+1}(\mathbf{A},i \leftarrow j) =
\mathrm{Net}(\mathbf{A},i \leftarrow j) + \delta_{ij},
\end{equation}
where $\delta_{ij}$ is the Kronecker delta,
\begin{equation}
\mathrm{Net}_{+1}(\mathbf{A},i \leftarrow \mathrm{all}) =
\mathrm{Net}(\mathbf{A},i \leftarrow \mathrm{all}) + 1,
\end{equation}
and
\begin{equation}
\mathrm{Net}_{+1}(\mathbf{A}, \mathrm{all} \leftarrow j) =
\mathrm{Net}(\mathbf{A},\mathrm{all} \leftarrow j) + 1.
\end{equation}

We have seen that net effects are well defined under the hypothesis that $\rho (\mathbf{A}) < 1$, but not otherwise. Consequently, we have to compute the spectral radius of $\mathbf{A}$, and confirm that its value lies below one to assure convergence.
Since computing the spectral radius can be very costly, it is often useful to apply the fact that if $\mathbf{A}$ has zero diagonal and $\mathbf{I}_n - \mathbf{A}$ is strictly diagonally dominant (SDD) by columns, then $\rho (\mathbf{A}) < 1$.
To see this, let $\mathbf{B} = [b_{ij}]$ be any $n \times n$ matrix, and let $C_j(\mathbf{B})$ denote the sum of the absolute values of the non-diagonal entries in the $j$-th column of $\mathbf{B}$.
Then $\mathbf{B}$ is SDD by columns if
\begin{equation}
C_j(\mathbf{B}) =
\sum_{i \not= j} |b_{ij}| <
|b_{jj}| \quad \mathrm{for \ all \ } j.
\end{equation}
Now, let us assume that $\mathbf{A}$ has zero diagonal and $\mathbf{I}_n - \mathbf{A}$ is SDD by columns, so that
\begin{equation}
C_j =
C_j(\mathbf{A}) =
C_j(\mathbf{I}_n - \mathbf{A}) =
\sum_{i \not= j} |a_{ij}| <
1 \quad \mathrm{for \ all \ } j.
\end{equation}
Then, by Gershgorin circle theorem (see e.g.~\cite{meyer2000}, p.498), every eigenvalue of $\mathbf{A}$ lies within at least one of the Gershgorin discs $\overline{D}(0,C_j)$, and hence $\rho (\mathbf{A}) < 1$.

The same conclusion applies if $\mathbf{A}$ has zero diagonal and $\mathbf{I}_n - \mathbf{A}$ is SDD by rows (the definition is entirely similar), since Gershgorin theorem is equally valid row-wise.

Classical centrality measures, such as Katz's or PageRank, also face the convergence issue, and solve it by rescaling the interaction matrix in a certain manner.
The idea is then to use net effects with respect to the rescaled interaction matrix as a proxy for (possibly undefined) net effects with respect to $\mathbf{A}$.
Rescaling the interaction matrix can be interpreted as a <<reweighting>> of the network, preserving the connection structure, the signs of the interactions, and the relative information of the weights.
There are three natural ways to achieve this for signed weighted directed networks, each addressed in one of the sections below.
Rescaling in one way or another depends on the research question; we will provide examples and intuition for ecological and social networks.

\subsubsection*{Global rescaling} \label{global}

The simplest rescaling procedure is to rescale $\mathbf{A}$ globally by a positive number $\alpha$, obtaining a rescaled interaction matrix $\alpha \mathbf{A} = [\alpha a_{ij}]$.
If we choose $\alpha$ such that $\rho (\alpha \mathbf{A}) = \alpha \, \rho (\mathbf{A}) < 1$, then
\begin{equation}
\mathrm{Net}(\alpha \mathbf{A}) =
- \mathbf{I}_n + (\mathbf{I}_n - \alpha \mathbf{A})^{-1} =
\sum_{\ell = 1}^{\infty} \alpha^\ell \mathbf{A}^\ell
\end{equation}
is well defined, and we have at our disposal all the net effect notions introduced earlier, but with respect to $\alpha \mathbf{A}$.
Notice that 
\begin{equation}
\mathrm{Net}(\alpha \mathbf{A},i \leftarrow j) =
\sum_{\ell = 1}^{\infty} \alpha^\ell \, [\mathbf{A}^\ell]_{ij}
\end{equation}
measures the net effect of node $j$ on node $i$, taking into account that the indirect effect of node $j$ on node $i$ along a walk of length $\ell$ has been <<attenuated>> by a factor of $\alpha^\ell$.

If $\mathbf{A}$ has zero diagonal and we choose $\alpha < 1 / (n - 1) M$, where
\begin{equation}
M \geq \max_{i, \, j} |a_{ij}|,
\end{equation}
then we can guarantee that $\rho (\alpha \mathbf{A}) < 1$, since in this case $\mathbf{I}_n - \alpha \mathbf{A}$ is SDD by columns:
\begin{equation}
C_j(\mathbf{I}_n - \alpha \mathbf{A}) =
\alpha \sum_{i \not= j} |a_{ij}| \leq
\alpha (n - 1) M < 1. \label{eq:kk}
\end{equation}

Interestingly, we can recover from our general framework classical measures of centrality defined for less general networks. For example, if the network is simple and directed, but unsigned and unweighted, then the vector of incoming net effects with respect to $\alpha \mathbf{A}$,
\begin{equation}
\mathrm{Net}(\alpha \mathbf{A},* \leftarrow \mathrm{all}) =
(- \mathbf{I}_n + (\mathbf{I}_n - \alpha \mathbf{A})^{-1}) \, \mathbf{1},
\end{equation}
is the \textbf{Katz centrality}~\cite{katz1953} vector with <<attenuation factor>> $\alpha$, and the vector of incoming plus one net effects with respect to $\alpha \mathbf{A}$,
\begin{equation}
\mathrm{Net}_{+1}(\alpha \mathbf{A},* \leftarrow \mathrm{all}) =
(\mathbf{I}_n - \alpha \mathbf{A})^{-1} \, \mathbf{1},
\end{equation}
is the \textbf{Hubbell centrality}~\cite{hubbell1965} vector with <<boundary condition>> $\mathbf{1}$.
Therefore, incoming net effects (respectively, incoming plus one net effects) with respect to $\alpha \mathbf{A}$ can be interpreted as a generalization of Katz centrality (respectively, Hubbell centrality) to signed weighted directed networks.

\subsubsection*{Column rescaling} \label{column}

Sometimes, a more sophisticated rescaling procedure is convenient.
We can take positive numbers $d_1, \ldots, d_n$, and rescale column $j$ of matrix $\mathbf{A}$ by $d_j^{-1}$.
This is equivalent to (locally) rescaling by $d_j^{-1}$ all links that go out from node $j$.
Moreover, we rescale all links (globally) by a positive number $\alpha$.
The resulting rescaled interaction matrix is $\alpha \mathbf{A} \mathbf{D}^{-1} = [\alpha a_{ij} d_j^{-1}]$, where $\mathbf{D} = \mathrm{diag}(d_1, \ldots, d_n)$.
Performing column rescaling in this manner, which may seem somewhat contrived at first reading, generalizes one of the standard ways to introduce PageRank (see e.g.~\cite{newman2018}, Section 7.1.4).

If $\mathbf{A}$ has zero diagonal, there are essentially two ways to choose $d_1, \ldots, d_n$ and $\alpha$ so that $\mathbf{I}_n - \alpha \mathbf{A} \mathbf{D}^{-1}$ is SDD by columns, and hence $\mathrm{Net}(\alpha \mathbf{A} \mathbf{D}^{-1})$ is well defined.
The first is to take $\alpha < 1$ and choose each $d_j$ so that
\begin{equation} \label{outdegree}
d_j \geq \sum_{i \not= j} |a_{ij}| = d_j^{\mathrm{\, out}}, 
\end{equation}
the weighted out-degree of node $j$.
Indeed,
\begin{equation}
C_j(\mathbf{I}_n - \alpha \mathbf{A} \mathbf{D}^{-1}) =
\alpha \, d_j^{-1} \sum_{i \not= j} |a_{ij}| \leq
\alpha < 1.
\end{equation}
Alternatively, we obtain the same result if $\alpha = 1$ and $d_j > d_j^{\mathrm{\, out}}$ for all $j$.

If the network is simple and directed, but unsigned and unweighted, $\alpha < 1$, and $d_j = d_j^{\mathrm{\, out}}$ is positive for all $j$ (i.e.~there are no <<dangling nodes>>), then the vector of incoming plus one net effects with respect to $\alpha \mathbf{A} \mathbf{D}^{-1}$,
\begin{equation}
\mathrm{Net}_{+1}(\alpha \mathbf{A} \mathbf{D}^{-1},* \leftarrow \mathrm{all}) =
(\mathbf{I}_n - \alpha \mathbf{A} \mathbf{D}^{-1})^{-1} \, \mathbf{1},
\end{equation}
is, up to a factor $(1 - \alpha) / n$, the classical \textbf{PageRank centrality} vector with <<teleportation parameter>> $\alpha$ and uniform <<teleportation distribution vector>> $\frac{1}{n} \, \mathbf{1}$ (see~\cite{gleich2015} for details).
Therefore, incoming plus one net effects with respect to $\alpha \mathbf{A} \mathbf{D}^{-1}$ can be interpreted as a generalization of PageRank centrality to signed weighted directed networks.

The second way to guarantee convergence is to take $\alpha < 1 / (n - 1)$ and choose each $d_j$ so that $d_j \geq |a_{ij}|$ for all~$i$, since in this case
\begin{equation}
C_j(\mathbf{I}_n - \alpha \mathbf{A} \mathbf{D}^{-1}) =
\alpha \, d_j^{-1} \sum_{i \not= j} |a_{ij}| \leq
\alpha (n - 1) < 1.
\end{equation}
Alternatively, we obtain the same result if $\alpha = 1 / (n - 1)$ and $d_j > |a_{ij}|$ for all~$i$ and $j$.

\subsubsection*{Row rescaling} \label{row}

Similarly, we can take positive numbers $d_1, \ldots, d_n$, and rescale row $i$ of matrix $\mathbf{A}$ by $d_i^{-1}$, which is equivalent to (locally) rescaling by $d_i^{-1}$ all links that come into node $i$.
As before, we also rescale all links (globally) by a positive number $\alpha$.
The resulting rescaled interaction matrix is $\alpha \mathbf{D}^{-1} \! \mathbf{A} = [\alpha d_i^{-1} a_{ij}]$, where $\mathbf{D}$ is defined as above.

Similarly to Section "Column rescaling", if $\mathbf{A}$ has zero diagonal, there are essentially two ways to choose $d_1, \ldots, d_n$, and $\alpha$ so that $\mathbf{I}_n - \alpha \mathbf{D}^{-1} \! \mathbf{A}$ is SDD by rows, and hence $\mathrm{Net}(\alpha \mathbf{D}^{-1} \! \mathbf{A})$ is well defined.
The first is to take $\alpha < 1$ and choose each $d_i$ so that
\begin{equation} \label{indegree}
d_i \geq \sum_{j \not= i} |a_{ij}| = d_i^{\mathrm{\, in}}, 
\end{equation}
the weighted in-degree of node $i$, and the second is to take $\alpha < 1 / (n - 1)$ and choose each $d_i$ so that $d_i \geq |a_{ij}|$ for all $j$.
Alternatively, we obtain the same result if $\alpha = 1$ and $d_i > d_i^{\mathrm{\, in}}$ for all $i$, or $\alpha = 1 / (n - 1)$ and $d_i > |a_{ij}|$ for all~$i$ and $j$.

Notice that the transposed matrix $\mathbf{A}^{\! T}$ represents the reversed of the original network (i.e.\ the direction of every link is reversed, but preserving its weight and sign).
If the network is simple and directed, but unsigned and unweighted, $\alpha < 1$, and $d_i = d_i^{\mathrm{\, in}}(\mathbf{A}) = d_i^{\mathrm{\, out}}(\mathbf{A}^{\! T})$ is positive for all $i$, then
\begin{equation}
[ \, \mathrm{Net}_{+1}(\alpha \mathbf{D}^{-1} \! \mathbf{A},\mathrm{all} \leftarrow *) \, ]^{\, T} =
[ \, \mathbf{1}^T (\mathbf{I}_n - \alpha \mathbf{D}^{-1} \! \mathbf{A})^{-1} \, ]^{\, T} =
(\mathbf{I}_n - \alpha \mathbf{A}^{\! T} \mathbf{D}^{-1})^{-1} \, \mathbf{1} =
\mathrm{Net}_{+1}(\alpha \mathbf{A}^{\! T} \mathbf{D}^{-1},* \leftarrow \mathrm{all}).
\end{equation}
Since $\mathrm{Net}_{+1}(\alpha \mathbf{A}^{\! T} \mathbf{D}^{-1},* \leftarrow \mathrm{all})$ is the PageRank centrality of the reversed network,
$\mathrm{Net}_{+1}(\alpha \mathbf{D}^{-1} \! \mathbf{A},\mathrm{all} \leftarrow *)$ is the \textbf{reverse PageRank centrality} of the original network~\cite{gleich2015}.
Therefore, outgoing plus one net effects with respect to $\alpha \mathbf{D}^{-1} \! \mathbf{A}$ can be interpreted as a generalization of reverse PageRank centrality to signed weighted directed networks.

Since all the measures of net effects introduced so far are defined in terms of matrix series, they can be approximated by their truncated versions.
For example, the $\mathrm{L}$-truncated vector of outgoing net effects with respect to $\alpha \mathbf{D}^{-1} \! \mathbf{A}$ is defined by
\begin{equation} \label{truncated}
\mathrm{Net}^{\mathrm{L}}(\alpha \mathbf{D}^{-1} \! \mathbf{A},\mathrm{all} \leftarrow *) =
\mathbf{1}^T \, [ \, \sum_{\ell = 1}^{\mathrm{L}} \alpha^\ell \, (\mathbf{D}^{-1} \mathbf{A})^\ell \, ].
\end{equation}
This definition will be used in Eq.~\eqref{ne}.

\subsection*{Net effects in ecological networks} \label{ecological}

Species interactions are a fundamental backbone for coexistence~\cite{kefi2020}. By encoding interactions in matrices, it is possible to measure the position and role of species in several ways, providing different ecological insights. Beyond direct interactions, indirect interactions in ecology occur when interactions are mediated through one or several intermediate species~\cite{wootton1994}. They form long pathways~\cite{puccia1991}, generating confounding causal effects between the species involved~\cite{bender1984,roughgarden1986}.
For example, <<apparent competition>> occurs when two species are predated by a shared natural enemy~\cite{miller2021}, while <<exploitative competition>> occurs through feeding on a shared resource~\cite{browett2023}. Another common type of indirect interaction takes place when the presence of a top predator initiates a <<trophic cascade>> by suppressing the abundance 
of an intermediate consumer, resulting in an increase in the abundance of lower trophic levels~\cite{montoya2009,lundgren2022}. One can also consider indirect interactions between fish and plants via dragonflies, whose larvae are eaten by fish but whose adults prey on plant pollinators~\cite{knight2005}. Consequently, indirect interactions can couple biomes, determine loops that control the stability of food webs, or impair our ability
\added{to manage natural resources~\cite{yodzis2001} or}
to control pests~\cite{loyn1983}.
When indirect interactions are considered, it is often unclear whether their net effect 
will be positive or negative. For example, direct parasitic interactions may ultimately result in a net commensal ---or even mutualistic--- relationship as indirect effects develop~\cite{price1986}.

Thus, net effects incorporate intricate interconnections generated by indirect interactions, leading to emergent community behavior that is different from isolated interaction pairs and motifs~\cite{loreau2020,aguade2026}. In ecology, net effects are also related to predictions and inferences regarding the community response to press perturbations  ---long-term disturbances to one or more species density or parameters~\cite{novak2016}. Since the interaction matrix $\mathbf{M}$ can be constructed in multiple ways (such as the alpha matrix, the beta matrix, or the community matrix), each choice yields a so-called <<net effects matrix>>~\cite{novak2016}, $-\mathbf{M}^{-1}$, which gives insights regarding different types of press perturbations. 
When $\mathbf{M}$ is defined as: (i) $\mathbf{M} = \mathbf{J}(\mathbf{x}^*)$ (the community matrix), (ii) $\mathbf{M} = \mathbf{A}$ (the interaction matrix), or (iii) $\mathbf{M} = \boldsymbol{\alpha}$ (the alpha matrix); then entry $(i,j)$ of $-\mathbf{M}^{-1}$ measures the response of the equilibrium abundance of species $i$ to a small press perturbation in (respectively) (i) the growth rate~\cite{yodzis1988,novak2016}, (ii) the intrinsic growth rate~\cite{roberts2004,novak2016}, or (iii) the carrying capacity~\cite{bender1984,novak2016} of species $j$. 

Critically, the connection of net effects as defined from $-\mathbf{M}^{-1}$ with net effects in our sense of <<overall effects through both direct and indirect pathways>>~\cite{wootton2002}, while explicitly suggested in many ecological articles~\cite{bender1984,yodzis1988,roberts2004,montoya2009,novak2016,higashi1995}, has just been made mathematically explicit in~\cite{zelnik2024} for the alpha matrix.
In this section, we complete the mathematical description of net effects matrices, and specify the conditions under which they yield meaningful insights into the different press perturbations.

To do so, we consider an ecological network of $n$ species whose dynamics are governed by generalized Lotka-Volterra (gLV) equations~\cite{case2000} of the form
\begin{equation} \label{gLV}
\dot{x}_i = x_i \left( r_i + \sum_{j=1}^n a_{ij} x_j \right), \quad i = 1,\ldots,n,
\end{equation}
where $x_i$ is the abundance of species $i$, $\dot{x}_i$ denotes its time derivative, which is the growth rate of species $i$, $r_i$ is the intrinsic growth rate of species $i$, and $a_{ij}$ measures the direct effect of species $j$ on the per capita growth rate of species $i$, i.e.~on $\dot{x}_i / x_i$.
The intraspecific coefficient $a_{ii}$ accounts for the self-regulation of species $i$, whereas the interspecific coefficients $a_{ij}$, for $j \not= i$, account for the pairwise interactions between species.
Depending on the signs of $a_{ij}$ and $a_{ji}$, these interactions can be predator-prey $(+ \, -)$, competitive $(- \, -)$, mutualistic $(+ \, +)$, etc.
The matrix $\mathbf{A} = [a_{ij}]$ is the interaction matrix of our ecological network~\cite{novak2016}.

As is common in theoretical ecology, we will assume that all species have negative self-regulation, i.e.~$a_{ii} < 0$ for all $i$ (see Section 5.1 of~\cite{novak2016} for a discussion of this topic).
We will also assume that Eqs.~\eqref{gLV} have a locally stable feasible equilibrium, i.e.~a vector $\mathbf{x}^* = (x_1^*,\ldots,x_n^*)^T$ of positive abundances such that $\mathbf{r} + \mathbf{A} \mathbf{x^*} = \mathbf{0}$, where $\mathbf{r} = (r_1,\ldots,r_n)^T$ is the vector of intrinsic growth rates, and all the eigenvalues of $\mathbf{J}(\mathbf{x}^*)$, the Jacobian matrix evaluated at $\mathbf{x}^*$, have negative real parts.
In particular, $\mathbf{J}(\mathbf{x}^*)$ is an invertible matrix, and since $\mathbf{J}(\mathbf{x}^*) = \mathbf{X}^* \mathbf{A}$, where $\mathbf{X}^* = \mathrm{diag}(x_1^*,\ldots,x_n^*)$, $\mathbf{A}$ is also an invertible matrix (see e.g.~\cite{case2000}, p.351 for details).

\subsubsection*{The alpha matrix} \label{alpha}

The alpha matrix~\cite{novak2016} is a row-rescaled version of the interaction matrix $\mathbf{A} = [a_{ij}]$, in which the direct effect of species $j$ on (the per capita growth rate of) species $i$ is measured relatively to the direct effect, in absolute value, of species $i$ on itself, i.e.~as $- a_{ij}/a_{ii}$.

To define it more formally, let $\mathbf{A}^\dagger = [a^\dagger_{ij}]$ be the off-diagonal part of $\mathbf{A}$, i.e.~the matrix of interspecific interactions, given by
\begin{equation}
a^\dagger_{ij} = a_{ij} \quad \mathrm{if \ } i \not= j \quad \mathrm{and} \quad 
a^\dagger_{ii} = 0 \quad \mathrm{for \ all \ } i.
\end{equation}
In general, we will use a dagger to denote the operation of resetting all the diagonal entries of a matrix to zero (nothing to do with the Hermitian adjoint in this context, it is simply that a dagger seemed appropriate for killing diagonal entries).
On the other hand, let $\mathbf{D} = \mathrm{diag}(-a_{11}, \ldots, -a_{nn})$ be the diagonal matrix of intraspecific interactions in absolute value, so that $\mathbf{A} = - \mathbf{D} + \mathbf{A}^\dagger$.
Since we are assuming that $- a_{ii} > 0$ for all $i$, $\mathbf{D}$ is an invertible matrix.
The alpha matrix is then defined by $\boldsymbol{\alpha} = \mathbf{D}^{-1} \mathbf{A} = - \mathbf{I}_n + \mathbf{D}^{-1} \mathbf{A}^\dagger = - \mathbf{I}_n + \boldsymbol{\alpha}^\dagger$, so that $\boldsymbol{\alpha} = [\alpha_{ij}] = [- a_{ij}/a_{ii}]$.

Since $\alpha_{ij} = - a_{ij}/a_{ii}$ measures the direct effect of species $j$ on species $i$ relative to the direct effect, in absolute value, that species $i$ has on itself, $\boldsymbol{\alpha}$ and $\boldsymbol{\alpha}^\dagger$ are non-dimensional, and therefore amenable to meaningful definitions of net effects ---recall the small aside after Eq.~\eqref{aside}.
Notice that $\mathbf{A}$ is not amenable since, no matter the dimension used to measure the population size (absolute abundance, density per area, density per volume), $a_{ij}$ cannot be non-dimensional~\cite{arditi2021}.

Now, let us assume that $r_i > 0$ for all $i$, and let $k_i = - r_i / a_{ii} > 0$ be the carrying capacity of species $i$.
In terms of the carrying capacities, Eqs.~\eqref{gLV} can be rewritten in the (up to a minus sign) classical form
\begin{equation} \label{KLV}
\dot{x}_i = \frac{r_i}{k_i} x_i \left( k_i + \sum_{j=1}^n \alpha_{ij} x_j \right), \quad i = 1,\ldots,n.
\end{equation}
Therefore, up to a minus sign, $\boldsymbol{\alpha}$ is the classical alpha matrix introduced by Levins~\cite{levins1968}.

Since we are assuming that $\mathbf{x}^*$ is a feasible equilibrium of Eqs.~\eqref{gLV}, it is also a feasible equilibrium of Eqs.~\eqref{KLV}, hence $\mathbf{k} + \boldsymbol{\alpha} \mathbf{x^*} = \mathbf{0}$, where $\mathbf{k} = (k_1,\ldots,k_n)^T$ is the vector of carrying capacities.
Moreover, since $\mathbf{A}$ is invertible, $\boldsymbol{\alpha}$ is also invertible, hence $\mathbf{x}^*$ is given by $\mathbf{x}^* = - \boldsymbol{\alpha}^{-1} \mathbf{k}$.
From this we deduce that 
\begin{equation}
    \partial x_i^* / \partial k_j = [- \boldsymbol{\alpha}^{-1}]_{ij}
\end{equation} 
represents the response of the equilibrium abundance of species $i$ to a small press perturbation in the carrying capacity of species $j$~\cite{bender1984,novak2016}.

The matrix $- \boldsymbol{\alpha}^{-1}$ is called the net effects matrix of $\boldsymbol{\alpha}$ in~\cite{novak2016}.
Its connection with net effects as presented here is given by~\cite{zelnik2024}:
\begin{equation} \label{eq:alphanet}
- \boldsymbol{\alpha}^{-1} = (\mathbf{I}_n - \boldsymbol{\alpha}^\dagger)^{-1} =
\sum_{\ell = 0}^{\infty} (\boldsymbol{\alpha}^\dagger)^\ell =
\mathrm{Net}_{+1}(\boldsymbol{\alpha}^\dagger).
\end{equation}
Importantly, Eq.~\eqref{eq:alphanet} is valid only under the hypothesis that $\rho (\boldsymbol{\alpha}^\dagger) < 1$.
This has led to the introduction in~\cite{zelnik2024} of the <<collectivity parameter>> $\phi = \rho (\boldsymbol{\alpha}^\dagger)$, as a measure of the degree to which net effects in an ecological community (in the sense of the net effects matrix $- \boldsymbol{\alpha}^{-1}$) emerge as infinite aggregates of indirect effects, and cease to do so when $\phi \geq 1$.

In the light of our presentation of net effects, a sufficient condition for $\mathrm{Net}_{+1}(\boldsymbol{\alpha}^\dagger)$ to be well defined is that $\mathbf{I}_n - \boldsymbol{\alpha}^\dagger$ or, equivalently, $\mathbf{A}$ itself, be SDD by rows, i.e.
\begin{equation}
- a_{ii} > \sum_{j \not= i} |a_{ij}| \quad \mathrm{for \ all \ } i.
\end{equation}
In the purely competitive case, this might be interpreted as saying that each species inhibits itself more than it is inhibited by \textit{all} the other species~\cite{strobeck1973}.
Note in passing that this implies that $\mathbf{A}$ is invertible (this is an immediate consequence of Gershgorin theorem, usually known as the L\'evy-Desplanques theorem).

\subsubsection*{The beta matrix} \label{beta}

The beta matrix~\cite{novak2016} is a column-rescaled version of the interaction matrix $\mathbf{A} = [a_{ij}]$, in which the direct effect of species $j$ on species $i$ is measured relatively to the direct effect, in absolute value, of species $j$ on itself, i.e.~as $- a_{ij}/a_{jj}$.

Similarly to the previous section, the beta matrix is then defined by $\boldsymbol{\beta} = \mathbf{A} \mathbf{D}^{-1} = - \mathbf{I}_n + \mathbf{A}^\dagger \mathbf{D}^{-1} = - \mathbf{I}_n + \boldsymbol{\beta}^\dagger$, so that $\boldsymbol{\beta} = [\beta_{ij}] = [-a_{ij}/a_{jj}]$.
Since $\beta_{ij} = - a_{ij}/a_{jj}$ measures the direct effect of species $j$ on species $i$ relative to the direct effect, in absolute value, that species $j$ has on itself, $\boldsymbol{\beta}$ and $\boldsymbol{\beta}^\dagger$ are non-dimensional.
Up to a minus sign, $\boldsymbol{\beta}$ is the beta matrix introduced by Vandermeer~\cite{vandermeer1975}.

Since we are assuming that $\mathbf{x}^*$ is a feasible equilibrium of Eqs.~\eqref{gLV}, we have that $\mathbf{r} + \mathbf{A} \mathbf{x^*} = \mathbf{0}$.
Moreover, since $\mathbf{A}$ is invertible, $\mathbf{x}^*$ is given by $\mathbf{x}^* = - \mathbf{A}^{-1} \mathbf{r}$.
From this we deduce that 
\begin{equation}
\partial x_i^* / \partial r_j = [- \mathbf{A}^{-1}]_{ij}    
\end{equation}
represents the response of the equilibrium abundance of species $i$ to a small press perturbation in the intrinsic growth rate of species $j$~\cite{roberts2004,novak2016}. However, even though the matrix $- \mathbf{A}^{-1}$ is called the net effects matrix of $\mathbf{A}$ in~\cite{novak2016},
it is not possible to relate this matrix to physically well-defined net effects, because $\mathbf{A}$ is not non-dimensional.
It is at this point that the beta matrix comes into play.

Let us introduce new <<population>> variables $\mathbf{u} = (u_1,\ldots,u_n)^T$, defined by $u_i = - a_{ii} x_i$ for all $i$, so that $\mathbf{u} = \mathbf{D} \mathbf{x}$, where $\mathbf{x} = (x_1,\ldots,x_n)^T$.
Since we are assuming that $- a_{ii} > 0$ for all $i$, this can be interpreted as a rescaling of the way in which we measure the abundance of each species separately.
Under this change of variables, Eqs.~\eqref{gLV} transform into a new set of gLV equations, in which the interaction matrix is precisely $\boldsymbol{\beta}$:
\begin{equation} \label{newgLV}
\dot{u}_i = u_i \left( r_i + \sum_{j=1}^n \beta_{ij} u_j \right), \quad i = 1,\ldots,n.
\end{equation}

Since $\mathbf{A}$ is invertible, $\boldsymbol{\beta}$ is also invertible, and the unique feasible equilibrium of Eqs.~\eqref{newgLV} is
\begin{equation}
\mathbf{u}^* = - \boldsymbol{\beta}^{-1} \mathbf{r} =
- \mathbf{D} \mathbf{A}^{-1} \mathbf{r} = \mathbf{D} \mathbf{x}^*.
\end{equation}
As before, $\partial u_i^* / \partial r_j = [- \boldsymbol{\beta}^{-1}]_{ij}$ represents the response of $u_i^*$ to a small press perturbation in $r_j$. The matrix $- \boldsymbol{\beta}^{-1}$ is the net effects matrix of $\boldsymbol{\beta}$ in the sense of~\cite{novak2016}. This time, the connection with net effects as defined in this work can be established, and is entirely analogous to that obtained for $- \boldsymbol{\alpha}^{-1}$ in the previous section:
\begin{equation}
- \boldsymbol{\beta}^{-1} = (\mathbf{I}_n - \boldsymbol{\beta}^\dagger)^{-1} =
\sum_{\ell = 0}^{\infty} (\boldsymbol{\beta}^\dagger)^\ell =
\mathrm{Net}_{+1}(\boldsymbol{\beta}^\dagger),
\end{equation}
which is valid only under the hypothesis that $\rho (\boldsymbol{\beta}^\dagger) < 1$.

A natural question that arises here is whether $\rho (\boldsymbol{\beta}^\dagger)$ is a new collectivity parameter of the ecological community, different from $\rho (\boldsymbol{\alpha}^\dagger)$.
That this is not the case, i.e.~that $\rho (\boldsymbol{\beta}^\dagger) = \rho (\boldsymbol{\alpha}^\dagger)$, can be deduced from the fact that similar matrices have the same eigenvalues (see e.g.~\cite{meyer2000}, p.508), and hence the same spectral radius:
\begin{equation}
\mathbf{D}^{-1} \boldsymbol{\beta}^\dagger \, \mathbf{D} =
\mathbf{D}^{-1} (\mathbf{A}^\dagger \, \mathbf{D}^{-1}) \, \mathbf{D} =
\mathbf{D}^{-1} \mathbf{A}^\dagger = \boldsymbol{\alpha}^\dagger.
\end{equation}

In the light of Section "Measures of net effects in signed weighted directed networks", a sufficient condition for $\mathrm{Net}_{+1}(\boldsymbol{\beta}^\dagger)$ to be well defined is that $\mathbf{I}_n - \boldsymbol{\beta}^\dagger$ or, equivalently, $\mathbf{A}$ itself, be SDD by columns, i.e.
\begin{equation}
- a_{jj} > \sum_{i \not= j} |a_{ij}| \quad \mathrm{for \ all \ } j.
\end{equation}
In the purely competitive case, this might be interpreted as saying that each species inhibits itself more than it inhibits \textit{all} the other species.
Whether each species inhibits itself more than it \textit{is inhibited by} or it \textit{inhibits} all the other species is an open question in coexistence theory.

Finally, we notice that a certain <<duality>> appears here: we can observe the system either from the alpha-matrix/carrying-capacities side, or from the beta-matrix/intrinsic-growth-rates side.
This duality, which deserves further research, is already explicit in the original article by Vandermeer~\cite{vandermeer1975}.

\subsubsection*{Net effects as structural predictors of species extinction} \label{gamma}

As a direct application of matrices of net effects to ecological networks, we report in this section that, under certain hypotheses, there is a strong correlation between species extinction and negative incoming net effect with respect to $\frac{1}{n} \boldsymbol{\beta}^\dagger$ (where $\boldsymbol{\beta}^\dagger$ is the off-diagonal part of the beta matrix).
Here, we no longer assume that our equations have a feasible equilibrium.

We claim that the vector $\mathrm{Net}(\frac{1}{n} \boldsymbol{\beta}^\dagger,* \leftarrow \mathrm{all})$ of incoming net effects with respect to $\frac{1}{n} \boldsymbol{\beta}^\dagger$ is a structural predictor of species extinction, in the sense that species that suffer a negative incoming net effect have a higher probability of becoming extinct under gLV dynamics or, equivalently, under replicator dynamics.

\newcommand{\methods}[2]{\hyperref[#1]{#2}}
To provide supporting evidence for our claim, we have conducted the computer simulations described in the "Methods" Section.
After completing the integration procedure described there, we count the number of extinct species, check whether the species with the minimum incoming net effect has become extinct, and calculate the proportion of species with negative incoming net effect that have become extinct.
\begin{figure}[h]
    \centering
    \includegraphics[width=\columnwidth]{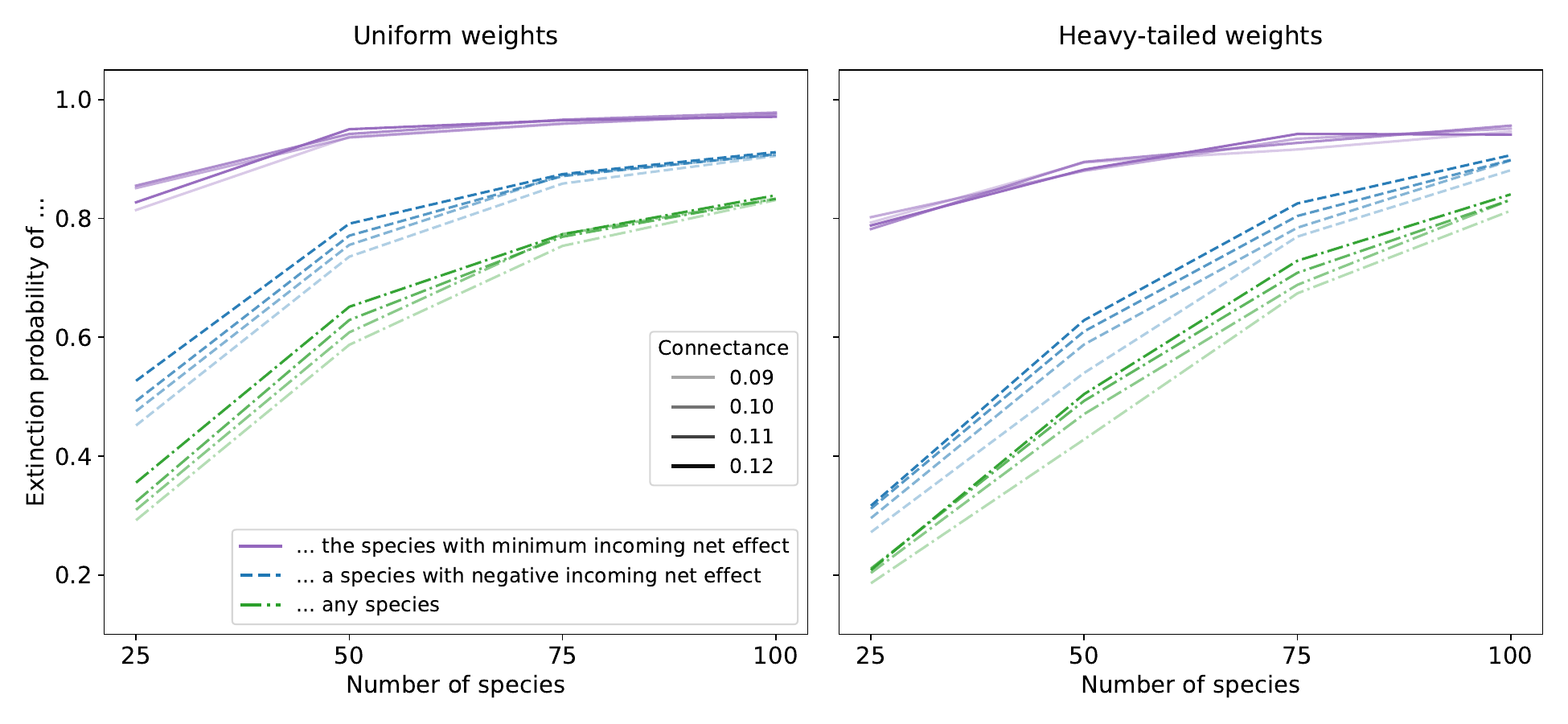}
    \caption{Average extinction probability, in random ecological networks with different numbers of species, and connectances, of the species with minimum incoming net effect, i.e. $ \mathrm{Net}\left(\frac{1}{n}\boldsymbol{\beta}^{\dagger}, * \leftarrow \mathrm{all} \right)$, a species with negative incoming net effect, and any species. For each pair of species number and connectance, we simulated 1000 \added{random interaction networks with weights drawn from a uniform distribution, left, or a Pareto distribution, right (see Methods for details and SI 1 for asymmetric sign distributions). The $y$-axis label text concludes with an ellipsis that maps directly onto the three alternative species categories defined in the legend.}}
    \label{fig:2}
\end{figure}
Fig.~\ref{fig:2} shows that species with negative values of incoming net effect become extinct with a higher probability than other species,
\added{both for a uniform and for a heavy-tailed weight distribution.}

Extinction is highly probable for the species with minimum incoming net effect.

We have also observed that if we consider the \textbf{mean net effect} with respect to $\frac{1}{n} \boldsymbol{\beta}^\dagger$, i.e. 
\begin{equation}
\langle \mathrm{Net}(\frac{1}{n} \boldsymbol{\beta}^\dagger) \rangle =
\frac{1}{n^2} \sum_{i,j = 1}^n \, [ \, \mathrm{Net}(\frac{1}{n} \boldsymbol{\beta}^\dagger) \, ]_{ij},
\end{equation}
and recalculate it along the process described above for the matrix that results every time a species becomes extinct, and the corresponding row and column are deleted, we obtain a function that is increasing on average.
This suggests that an <<optimization principle>> might be at work in the evolution of species abundances under gLV dynamics.

\subsection*{Net effects in social networks} \label{social}

In this section, we review some of the proposals of centrality or net effects measures for signed networks that have previously appeared in the social networks literature. 
We then embed these measures within our framework and test them against our measures of net effects in a standard example.

Works like~\cite{everett2014,liu2020,bonacich2004} illustrate their analyses with data from Sampson's well-known study of monk relationships in a monastery~\cite{sampson1968}, a key example for signed social networks, much like Zachary's karate club for community detection~\cite{girvan2002}.
Based on the monks' relationships, our task is to find a measure that predicts the four monks who were expelled: Basil, Elias, Simplicius, and Gregory~\cite{bonacich2004}.

The first proposed measure is the status score introduced in~\cite{bonacich2004}, based on classical eigenvector centrality.
As already noted in~\cite{everett2014}, this definition turns out to be problematic, since a matrix with positive and negative entries does not need to have a dominant eigenvalue.

The second proposal is PN centrality, introduced in~\cite{everett2014}.
The authors assume that their network is simple and signed, but neither weighted nor directed, i.e.~$\mathbf{A}$ is symmetric and $a_{ij} = 0, 1$ or $-1$ ($a_{ii} = 0$ for all $i$).
In order to define the PN centrality of $\mathbf{A}$, they proceed to asymmetrically rescale the positive and negative links with no clearly stated justification, making the interpretation of this measure rather opaque.
Specifically, their definition, translated into the language of Section "Measures of net effects in signed weighted directed networks", is
\begin{equation}
\mathrm{PN} (\mathbf{A}) =
\mathrm{Net}_{+1}(\frac{1}{n-1} \mathbf{B},* \leftarrow \mathrm{all}),
\end{equation}
where $\mathbf{B} = [b_{ij}]$ is the matrix obtained from $\mathbf{A}$ by rescaling the positive entries by $1/2$, i.e.~$b_{ij} = 1/2$ if $a_{ij} = 1$, and $b_{ij} = a_{ij}$ otherwise.
Under the hypothesis that the positive degree of each node is positive, i.e.~that every column of $\mathbf{B}$ contains at least one $1/2$ entry, we can guarantee that $\rho(\frac{1}{n-1} \mathbf{B}) < 1$, since in this case the matrix
$\mathbf{I}_n - \frac{1}{n-1} \mathbf{B}$ is SDD by columns:
\begin{equation}
C_j (\mathbf{I}_n - \frac{1}{n-1} \mathbf{B}) =
\frac{1}{n-1} \sum_{i \not= j} |b_{ij}| < 1.
\end{equation}

In our opinion, the good qualities attributed to $\mathrm{PN} (\mathbf{A})$ in the analysis of the Sampson monastery provided in~\cite{everett2014} are equally satisfied by
$\mathrm{Net}_{+1}(\frac{1}{n} \mathbf{A},* \leftarrow \mathrm{all})$.
The study also contains in and out versions of PN centrality for the directed case, but we have not been able to incorporate them into the framework presented here.

\begin{figure}[h]
\centering
\includegraphics[width = 0.5\linewidth]{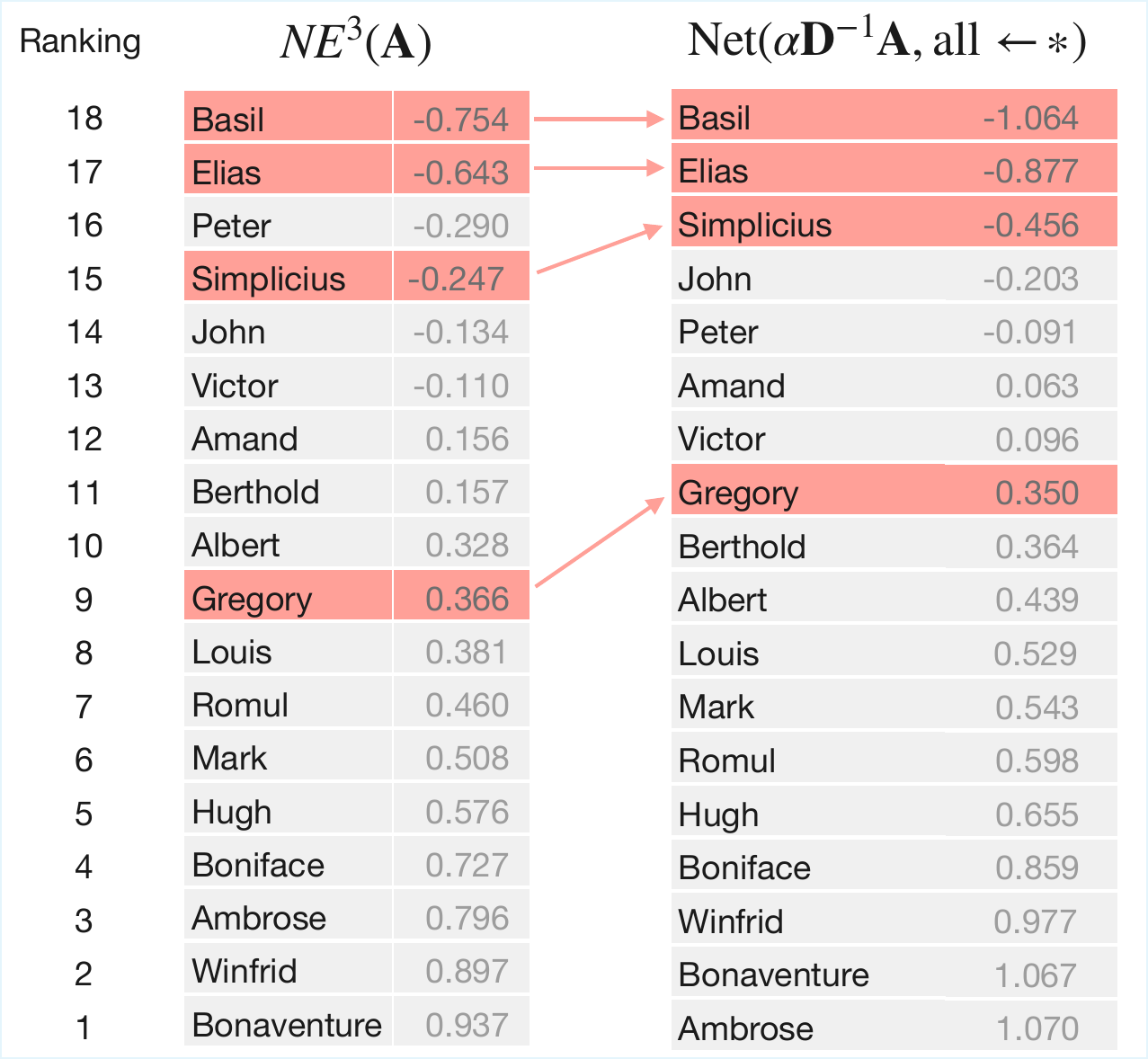}
\caption{Comparison of the monks' ranking according to the net effect truncated at walks of length 3 calculated in~\cite{liu2020}, left column, with the ranking when all walks are taken into account, right column. The right column better captures the actual ranking based on the finally expelled monks (in red). \added{See SI 2 for a null model analysis of the ranks.}}
\label{fig:3}
\end{figure}

The third proposal is the alternative definition of <<net effects>> introduced in~\cite{liu2020}
\added{for social networks, which builds upon an earlier formulation introduced in~\cite{liu2010} for food webs}.
In the most general case, the authors assume that their network is simple, signed and directed, but not weighted, i.e.~$a_{ij} = 0, 1$ or $-1$ ($a_{ii} = 0$ for all $i$), but $\mathbf{A}$ is not necessarily symmetric.
After a careful translation of the authors' methodology into matrix language (including a translation from row notation $i \to j$ into column notation $i \leftarrow j$), their definition of the vector $\mathrm{NE}^{\mathrm{L}}(\mathbf{A})$, whose $j$-th component is the net effect that node $j$ has on the whole network up to $\mathrm{L}$ steps with respect to the <<weighting function>> $g(\ell) = \alpha^{\ell}$, is
\begin{equation} \label{ne}
\mathrm{NE}^{\mathrm{L}}(\mathbf{A}) =  
\mathbf{1}^T \, [ \, \sum_{\ell = 1}^{\mathrm{L}} \alpha^\ell \, (\mathbf{D}^{-1} \mathbf{A})^\ell \, ] =
\mathrm{Net}^{\mathrm{L}}(\alpha \mathbf{D}^{-1} \! \mathbf{A},\mathrm{all} \leftarrow *),
\end{equation}
where $\mathbf{D} = \mathrm{diag}(d_1^{\mathrm{\, in}},\ldots,d_n^{\mathrm{\, in}})$, $d_i^{\mathrm{\, in}}$ is the total in-degree of node $i$ as in Eq.~\eqref{indegree}, and the definition of
$\mathrm{Net}^{\mathrm{L}}(\alpha \mathbf{D}^{-1} \! \mathbf{A},\mathrm{all} \leftarrow *)$ appeared in Eq.~\eqref{truncated}.
Since the authors consider $g$ to be identically 1, we can set $\alpha = 1$ in Eq.~\eqref{ne}. 
Hence, in the light of Section "Row rescaling",
$\mathrm{NE}^{\mathrm{L}}(\mathbf{A})$ can be interpreted as a truncated version of reverse PageRank centrality for signed directed simple networks.
In Fig.~\ref{fig:3}, we show the net effect
$\mathrm{NE}^3(\mathbf{A})$ up to $\mathrm{L} = 3$ steps calculated on the Sampson monastery like-dislike relationship network as it appears in~\cite{liu2020}, and compare it with the full version of
$\mathrm{Net} (\alpha \mathbf{D}^{-1} \! \mathbf{A},\mathrm{all} \leftarrow *)$
for $\alpha = 0.99 < 1$.
The latter seems to reflect the monks ranking more realistically in that the four monks who were expelled are worse positioned.
\added{In fact, achieving the exact same ranking for $\mathrm{NE}^{\mathrm{L}}(\mathbf{A})$ as the non-truncated $\mathrm{Net} (\alpha \mathbf{D}^{-1} \! \mathbf{A},\mathrm{all} \leftarrow *)$ formulation requires extending the path length to $\mathrm{L} = 13$, even if slight divergences in individual scores persist. The number of steps $\mathrm{L}$ necessary to retain precise structural information appears to be inherently network-dependent. This high variability strongly prompts the use of the full, non-truncated version over arbitrary truncation choices.}

\subsubsection*{Social replicator dynamics}

\begin{figure*}[ht]
\centering
\includegraphics[width =0.9\linewidth]{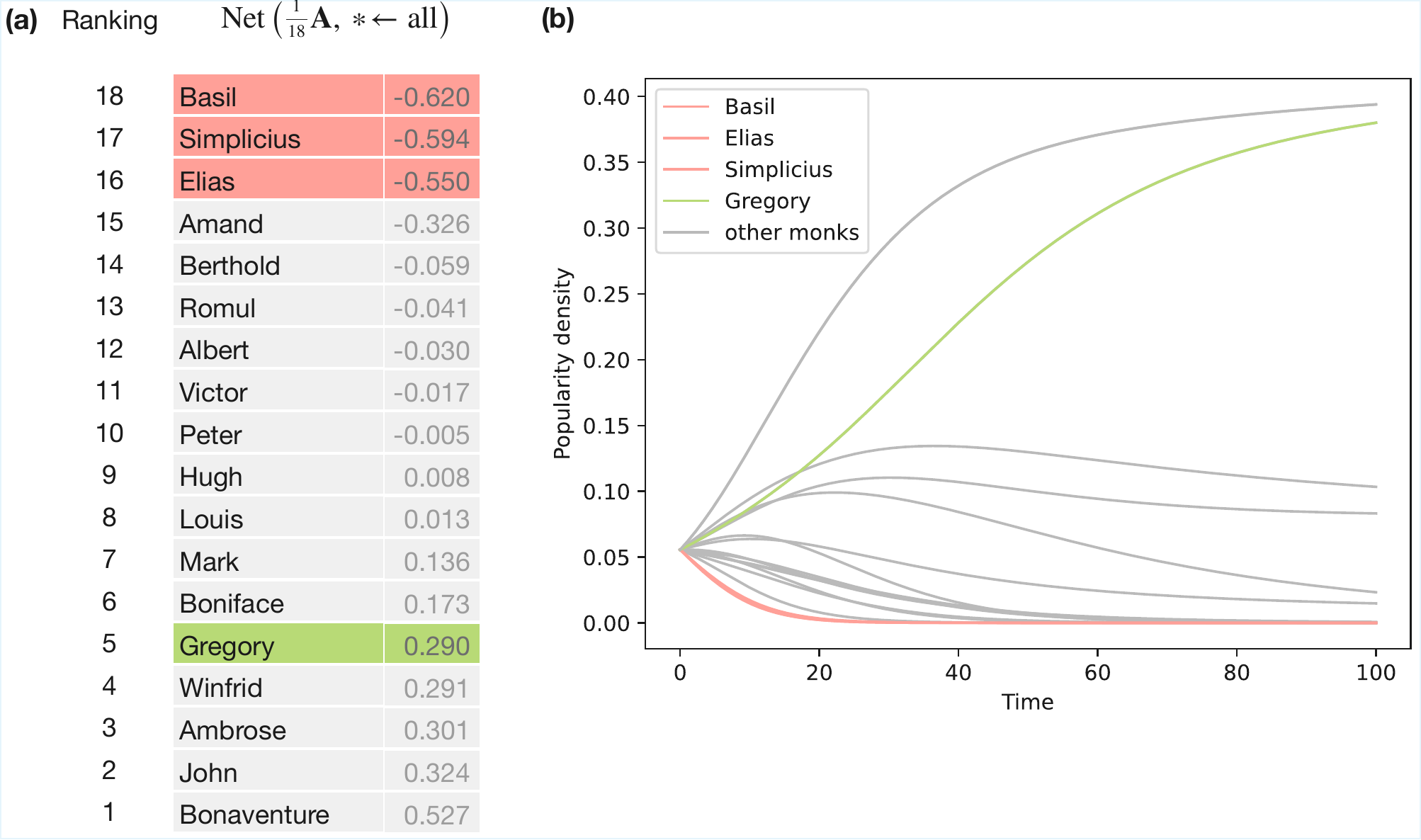}
\caption{\textbf{(a)} Ranking of the monks given by their incoming net effects and \textbf{(b)} evolution of their popularity as modeled by replicator dynamics.}
\label{fig:4}
\end{figure*}

In this final section, we apply the methods from Section "Net effects as structural predictors of species extinction" to the social network of the $18$ monks of Sampson's monastery.
We have taken from the Ucinet software~\cite{borgatti2002} the four matrices LK3, ES, IN and PR of positive interactions (<<liking>> at time period three, <<esteem>>, <<positive influence>>, and <<praise>>) and the corresponding four matrices DLK, DES, NIN and NPR of negative interactions (<<disliking>>, <<disesteem>>, <<negative influence>>, and <<blame>>), and we have formed the full matrix of interactions
\begin{equation}
\mathbf{A} = \mathrm{sgn}(\mathrm{LK3}^T + \mathrm{ES}^T + \mathrm{IN}^T + \mathrm{PR}^T) -
\mathrm{sgn}(\mathrm{DLK}^T + \mathrm{DES}^T + \mathrm{NIN}^T + \mathrm{NPR}^T),
\end{equation}
where the matrices have been transposed to comply with our column notation, and $\mathrm{sgn}(\mathbf{B})$ denotes the matrix obtained from a matrix $\mathbf{B}$ by applying the sign function entry-wise (i.e.~we have dichotomized the positive and negative interactions separately, as in~\cite{everett2014}, Table 7).
Next, we have computed the vector
$\mathrm{Net}(\frac{1}{18} \mathbf{A},* \leftarrow \mathrm{all})$
of incoming net effects with respect to $\frac{1}{18} \mathbf{A}$,
whose values can be seen in Fig.~\ref{fig:4}a.
Of the four monks that were expelled from the monastery, this measure captures Basil, Elias and Simplicius as the three monks with the lowest (negative) incoming net effect, but fails to capture Gregory, who has a high (positive) incoming net effect.
Finally, identifying fitness with popularity and starting in the center of the simplex, we have simulated the replicator dynamics with interaction matrix $\mathbf{A}$.
In agreement with the incoming net effect scores, Basil, Elias, and Simplicius are the first three monks whose popularity dwindles, while Gregory is the only one among them who ends up with a positive popularity score, as represented in Fig.~\ref{fig:4}b.
\added{The apparent discrepancy between Gregory's high position in the ranking (or high popularity density) and the fact that he was expelled from the monastery can be resolved by turning to Sampson's own analysis (see~\cite{sampson1968}, p. 373). He identifies Basil, Elias, and Simplicius (the <<Outcasts>>) as the three least popular monks, but characterizes Gregory (the leader of the <<Young Turks>>; see also~\cite{everett2014}) as an influential actor. Gregory was not expelled because of his unpopularity, but because the monastery leadership wanted to neutralize his influence. Therefore, Sampson's analysis provides a direct empirical validation of our model's capacity to correctly calculate net effects.}

\section*{Discussion}

In this paper, we have presented a mathematical description of matrices of net effects. Our unified framework computes net effects in signed, weighted, and directed networks, and organizes their variants into a coherent taxonomy. We deliberately use the plural ---matrices of net effects--- because allowing signed, weighted, and directed links yields multiple non-equivalent constructions (e.g.~by rescaling scheme and by incoming vs.~outgoing aggregation). We have (i) defined Net matrices that incorporate direct and indirect effects across paths of all lengths; (ii) derived the conditions under which net effects can be defined ---non-dimensional interaction matrices with spectral radius less than one---; and (iii) introduced three rescaling schemes ---global, column-wise, and row-wise--- that guarantee convergence while preserving the information of signs, directionality, and relative weights. Together, these elements permit incoming (<<node $\leftarrow$ network>>) and outgoing (<<network $\leftarrow$ node>>) net-effect summaries that address complementary questions.

To the best of our knowledge, this work goes beyond what has been studied so far. Our framework unifies and extends several classical measures. Incoming net effects with global rescaling generalize Katz centrality (and, in its plus-one form, Hubbell's) to signed, weighted, and directed networks; column-rescaled plus-one net effects recover PageRank, and row-rescaled plus-one net effects yield its reverse PageRank counterpart. We show that other measures that have appeared in the social network literature, like PN centrality and truncated <<net effect>> scores, can be interpreted as particular or truncated instances of our 
$\mathrm{Net}$ / $\mathrm{Net}_{+1}$ series under appropriate rescalings.
In ecology, the formal investigation of indirect effects via the Neumann series was pioneered by~\cite{hannon1973} and extensively developed by Patten and colleagues (see e.g.~\cite{higashi1989}). While these foundational analyses introduced the conceptual framework of network propagation, they restricted their attention to energy flows within unsigned networks. More recently, this methodological lineage has been revitalized by contemporary contributions such as~\cite{zelnik2024,aguade2026}. However, their framework focuses specifically on the $\boldsymbol{\alpha}$ matrix, leaving a gap regarding generalizations to signed, weighted, and directed networks, as well as alternative rescaling formulations. Likewise, a <<contribution of indirect effects>> is quantified in~\cite{guimaraes2017} for binary networks. In contrast, we connect press-perturbation interpretations with alpha and beta matrices, and make explicit when the ambiguous <<net effects>> of ecological literature coincide with the mathematically well-defined net effects of these matrices ($\boldsymbol{\alpha}^\dagger$, $\boldsymbol{\beta}^\dagger$), thereby clarifying earlier heuristic links.

Our applications illustrate a more complete perspective than traditional unsigned measures. In ecological simulations, the incoming net effect with respect to $\boldsymbol{\beta}^\dagger$ acts as a structural predictor of extinction risk: species with more negative incoming net effect go extinct more often, and the species with the minimum incoming value is frequently the first to vanish. In a signed social network, full net-effects measures (rather than their truncated versions) better align the ranking with observed outcomes in the Sampson monastery example, and simple social replicator dynamics mirrors those predictions. These results provide a powerful lens for ecosystem conservation and underscore the relevance of net effects for understanding phenomena such as influence propagation and power dynamics.
\added{Our matrix-based framework complements established methodologies like qualitative loop analysis~\cite{puccia1991} and the species uniqueness analysis of~\cite{liu2025}. Loop analysis evaluates qualitative signed digraphs to determine feedback stability under press perturbations. In contrast, our approach requires quantified interaction weights to calculate the exact numerical magnitudes of directional net effects. Furthermore, while~\cite{liu2025} analyze interaction profiles to identify structurally irreplaceable nodes, our taxonomy aggregates these multi-step pathways into a convergent algebraic series that maps the influence of each node.}

Importantly, our results do not rest on the assumption of underlying Lotka-Volterra or replicator dynamics; the approach is pertinent to any differentiable model in which the interaction matrix encodes direct effects. When our analysis assumes a simple network (with zero diagonal), we make this explicit and use daggered matrices ($\dagger$) to separate inter- from intra-specific terms. Because units matter, the path-sum interpretation is physically meaningful only for non-dimensional matrices; our global/row/column rescalings provide interpretable routes to non-dimensionality. In an ecological context, note also that the indirect effects captured here are path-mediated via pairwise links; they should not be confused with trait-mediated interactions or interaction modifications that alter the direct links themselves.

The inclusion of sign, weight, and direction in links makes our framework broadly applicable to real-world networks where relationships are inherently heterogeneous. However, challenges remain. First, empirical links may contain sign/weight errors, and then sensitivity analyses (perturbing entries and averaging, comparing to truncated versions) are warranted. Second, the choice among global vs.~column vs.~row rescaling is problem-dependent (whether attenuating by walk length vs.~normalizing by in- or out-strength) and can change values and rankings; reporting both the choice and its rationale is advised. \added{Just as these mathematical choices are context-dependent, the underlying structure of the networks is highly domain-specific \cite{alon2002}. For example, classical food webs are characterized by strict $(+ \, -)$ pairings \cite{neutel2002}, with non-trophic interactions coupled in non-random patterns \cite{kefi2016}, whereas social networks present high associativity and dense clique formation \cite{Newman2003}. Systematically comparing how these distinct, domain-specific structural signatures influence and constrain the distribution of net effects is a critical next step. This is especially urgent in ecology, where our} applications were tested on random interactions. Given the growing evidence that non-random structures play a critical role in species maintenance~\cite{garcia2023,gellner2016}, studying real-world interaction networks and controlled synthetic structures will reveal how indirect interactions shape coexistence and community assembly~\cite{calleja2026general,cosmo2023} once empirical weights and signs are fully incorporated.

Integrating this framework with empirical datasets that contain information on signs, weights, and directions of links may challenge the views obtained from the unsigned, unweighted, or undirected versions of these networks. Since awareness of the importance of conducting long-term/longitudinal studies brings the opportunity to obtain these three link properties, the time is ripe to revisit results that combine network structure with dynamics for a wide variety of systems. We provide two practical tools: (i) rescaling choice and (ii) incoming vs.~outgoing aggregation, to let researchers match the metric to their question, while preserving interpretability across domains.

\section*{Methods}

\subsection*{Simulations of species extinctions with replicator dynamics}

As in Section "The alpha matrix", let us assume that $r_i > 0$ for all $i$, and that $k_i = - r_i / a_{ii} > 0$ is the carrying capacity of species $i$.
If we return to Eqs.~\eqref{KLV}, and change variables to relative yields $y_i = x_i / k_i$, we obtain
\begin{equation} \label{yield1}
\dot{y}_i =  r_i y_i \left( 1 + \sum_{j=1}^n \frac{\alpha_{ij} k_j}{k_i} y_j \right), \quad i = 1,\ldots,n.
\end{equation}

As is common in theoretical ecology, we will further assume that all intrinsic growth rates are identical, i.e.~$r_i = r$ for all $i$ (see~\cite{song2018} for a discussion of this topic).
Under this hypothesis,
\begin{equation}
\frac{\alpha_{ij} k_j}{k_i} =
(- \frac{a_{ij}}{a_{ii}})(- \frac{r}{a_{jj}})(- \frac{a_{ii}}{r}) =
- \frac{a_{ij}}{a_{jj}} = \beta_{ij},
\end{equation}
so that Eqs.~\eqref{yield1} reduce to
\begin{equation} \label{yield2}
\dot{y}_i =  r y_i \left( 1 + \sum_{j=1}^n \beta_{ij} y_j \right), \quad i = 1,\ldots,n.
\end{equation}
Moreover, if we rescale time by letting $\tau = r t$, we can eliminate $r$ from Eqs.~\eqref{yield2}.

If we finally change variables to $z_i = y_i (y_1 + \cdots + y_n)^{-1}$ and apply~\cite{hofbauer1998}, Exercise 7.5.2, we arrive at the replicator equations
\begin{equation} \label{replicator}
\dot{z}_i = z_i ((\boldsymbol{\beta} \mathbf{z})_i - \mathbf{z}^T \! \boldsymbol{\beta} \mathbf{z}), \quad i = 1,\ldots,n,
\end{equation}
on the $n$-dimensional simplex $S_n$, where $\mathbf{z} = (z_1, \ldots, z_n)^T$ and $(\boldsymbol{\beta} \mathbf{z})_i$ denotes the $i$-th component of vector $\boldsymbol{\beta} \mathbf{z}$.
These equations, which are the ones we will use in our simulations, have the advantage of confining the dynamics to a bounded phase space.

Let us finally assume that $-a_{jj} > |a_{ij}|$ for all $i$ and $j$, $i \not= j$.
In the purely competitive case, this might be interpreted as saying that each species inhibits itself more than it inhibits \textit{any} of the other species (see~\cite{chesson2000}, p.345).
Under this hypothesis, $|b^\dagger_{ij}| < 1$ for all $i$ and $j$, so that we can take $M = 1$ and $\alpha < 1 / (n-1)$ to ensure that $\mathrm{Net}(\alpha \boldsymbol{\beta}^\dagger)$ is well defined (see Section "Global rescaling").
We have opted for $\alpha = 1 / n$ as a sensible choice.
Note that the global rescaling parameter $\alpha$ should not be confused with the alpha matrix $\boldsymbol{\alpha}$ (both notations are classical).

In order to test our claim, we have averaged over 1000 realizations of the following process for communities of different sizes and network connectances.
We begin by selecting a connected matrix $\boldsymbol{\beta} = -\mathbf{I}_n + \boldsymbol{\beta}^\dagger$ with off-diagonal entries randomly chosen from a uniform distribution in the interval $(-1,1)$, once a proportion $c$ of pairs $(i,j)$ and $(j,i)$, with $j \not= i$, have been selected for non-zero interaction.
\added{This is the same as an Erdős–Réyi graph with connectance $c$ and  uniformly random weights in $(-1,1)$.}
Then we compute $\mathrm{Net}(\frac{1}{n} \boldsymbol{\beta}^\dagger,* \leftarrow \mathrm{all})$, and store the species with minimum (negative) incoming net effect, as well as those with negative incoming net effect.
Next, we sample an initial condition uniformly from the simplex $S_n$, and numerically integrate the replicator Eqs.~\eqref{replicator} using the Runge–Kutta method of order 4 until the system stabilizes.
At each time step, we check for species whose densities have fallen below $10^{-5}/n$, which is the threshold density we have chosen for a species to be considered extinct, and reset their densities to zero, as well as the corresponding rows and columns of matrix $\boldsymbol{\beta}$.
For the sake of numerical stability, we normalize the resulting vector of densities to replace it within the simplex. \added{To represent heterogeneous ecological interactions, we also sampled absolute weights from a heavy-tailed Pareto distribution with a shape parameter $\alpha = 2.5$ and a scale factor of $0.5$. Crucially, the underlying connectance was kept strictly identical to the baseline uniform case, and signs were subsequently assigned with equal probability to ensure a controlled structural comparison.}
\newpage
\bibliography{references}

\section*{Acknowledgements}
The authors are grateful to the editor and the anonymous reviewers for their constructive comments, which have greatly improved the manuscript.
They also wish to thank Oscar Godoy, Fernando Diaz-Diaz, as well as Victor Maull and the ICREA-Complex Systems Lab members for insightful conversations.

\section*{Author contributions statement}

CGA: Conceptualization (supporting); Formal analysis (lead); Methodology (lead); Software (equal); Writing - original draft (equal); Writing - review and editing (equal).\\
VCS: Conceptualization (lead); Formal analysis (supporting); Methodology (supporting); Software (equal); Supervision (lead); Visualization (lead); Writing - original draft (equal); Writing - review and editing (equal).

\section*{Data and code availability}
Notebooks to reproduce all the results and ready-to-go functions of all the measures of net effect are available online at the Zenodo repository  \href{https://zenodo.org/records/17288129}{https://zenodo.org/records/17288129}. The datasets analysed during the current study (the eight matrices of Sampson's monastery) are also available in the Zenodo repository \href{https://zenodo.org/records/17288129}{https://zenodo.org/records/17288129}.

\section*{Additional information}

\textbf{Competing interests} 
The authors declare no competing interests.\\
\textbf{Funding} 
CGA acknowledges financial support from the Ministerio de Ciencia, Innovación y Universidades (grant PID2023-147734NB-I00). VCS acknowledges financial support from the Ministerio de Ciencia e Innovación (grant PID2021-127607OB-I00), and NERC Pushing the Frontiers grant (NE/
X013766/1).

\end{document}


\title{Supplementary Information of a unified framework for net effects in signed social and ecological networks}

\title{A unified framework for net effects in signed social and ecological networks \\ Supplementary information}

\author{ Carlos G\'omez-Ambrosi\,\orcidlink{0000-0001-7951-380X}}
\affiliation{Department of Mathematics, University of Zaragoza, 50009 Zaragoza, Spain}
\affiliation{GOTHAM lab, Institute for Biocomputation and Physics of Complex Systems (BIFI), University of Zaragoza, 50018 Zaragoza, Spain}
\author{Violeta Calleja-Solanas\,\orcidlink{0000-0001-7917-8984}}
\email{violeta.calleja@biology.ox.ac.uk}
\affiliation{Department of Biology, University of Oxford, Life and Mind Building, Oxford OX1 3PS, United Kingdom}
\affiliation{Department of Ecology and Evolution, Doñana Biological Station (EBD-CSIC), 41092 Seville, Spain}

\maketitle
\onecolumngrid

\section*{Supplementary Note 1: Figure 2 with asymmetric sign distributions}

The positive and negative signs in the networks of Figure 2 were uniformly distributed. To explore possible biases emerging from the sign distributions, we have simulated networks with a heavy-tailed weight distribution,  but with a ratio of positive:negative signs of 2:3 (negative biased) and 3:2 (positive biased), instead of 1:1. \\

We find that the extinction probabilities increase in all cases, but there is still a tendency for the extinction probability of the species with the minimum incoming net effect to be the greatest among all the cases, as shown in Fig. S1:

\begin{figure}
    \includegraphics[width=\linewidth]{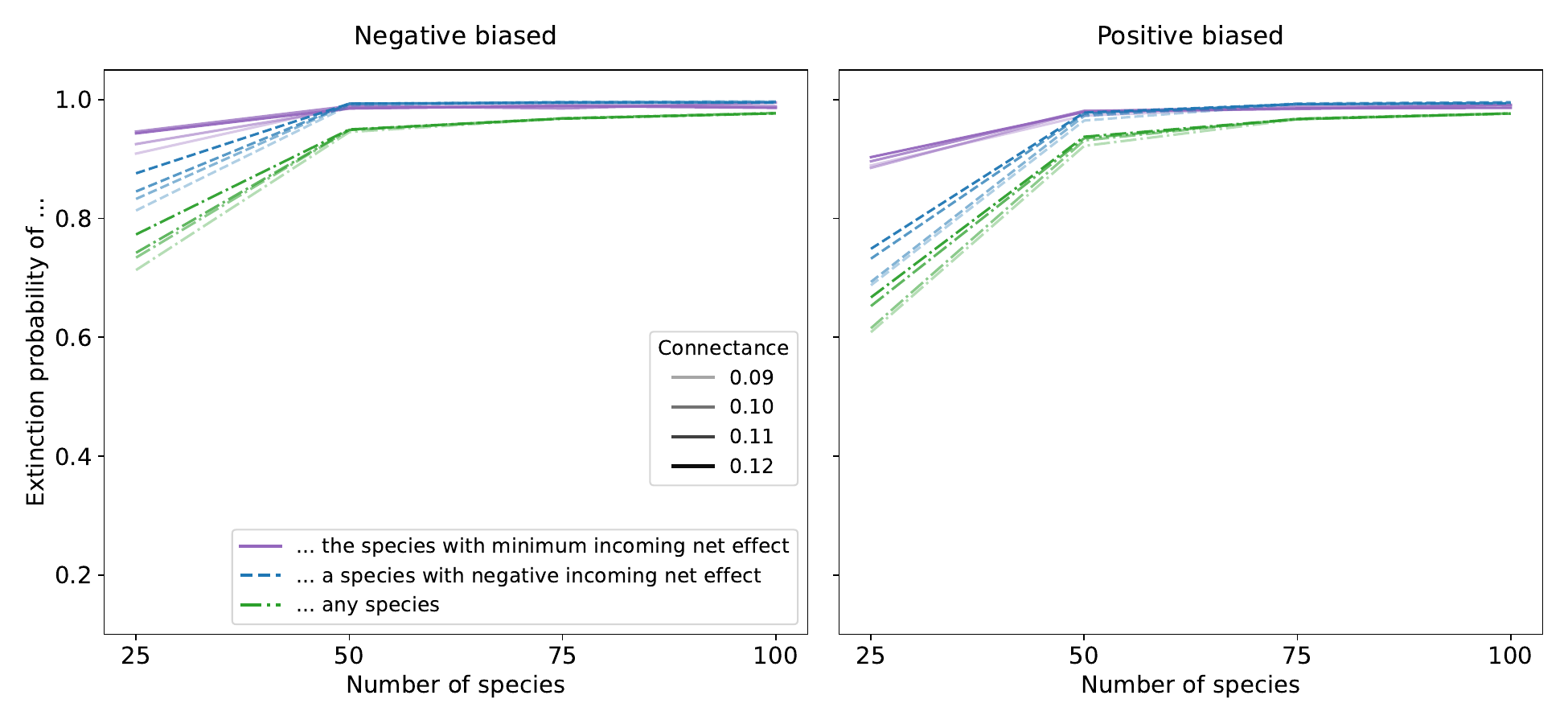}
    \caption{Average extinction probability, in random ecological networks with different numbers of species and connectances, of the species with minimum incoming net effect, i.e. $ \mathrm{Net}\left(\frac{1}{n}\boldsymbol{\beta}^{\dagger}, * \leftarrow \mathrm{all} \right)$, a species with negative incoming net effect, and any species. For each species number, we simulated 1000 random interaction networks with weights drawn from a heavy-tailed Pareto distribution (connectance = $0.12$, $\alpha = 2.5$, and scale factor $ = 0.5$). The $y$-axis label text concludes with an ellipsis that maps directly onto the three alternative species categories defined in the legend. Negative signs were drawn at random with probability $0.6$ ("Negative biased", left panel) and $0.4$ ("Positive biased", right panel).}
\end{figure}

\newpage

\section*{Supplementary Note 2: Null model analyses} 

To assess the statistical significance of the rankings presented in the main Fig. 3, we evaluated the empirical net effect rankings against two distinct network null models. These models were designed to disentangle the influence of local node constraints (degree and strength sequences) from the specific arrangement of positive and negative interactions.\\

\textbf{Null model 1.} We first implemented the simulated annealing algorithm proposed by Milisav et al. (2025) \cite{milisav2025}, which randomizes network topology while strictly preserving the exact weighted degree sequences for both positive and negative links. Because the empirical monks network is small ($N=18$) and dense, these rigid constraints severely restrict the accessible state space. Consequently, there are very few mathematical ways to rewire the edges without violating the algorithm's strict rules. When evaluating 100 iterations of this null model, the rank histograms for some monks (Simplicius, Romul, and Elias) exhibit biased distributions as expected, although the distributions are binomial. The histograms for the other monks present a less clear bias, as shown in Fig. S2. \\

To confirm that the results under the previous null model were biased for our small network size ($N=18$), rather than an inherent bias of the net effects metric, we evaluated the framework on larger synthetic networks. We generated 100-node directed networks featuring heavy-tailed interaction weights (connectance = 0.12; weights drawn from a Pareto distribution with scale = 0.5 and shape = 2.5). We then subjected these networks to the same strict simulated annealing algorithm, preserving the exact in/out degree and strength sequences. In this case, the $N=100$ networks allowed for sufficient variation while obeying the algorithm's constraints. Across the randomizations, the rank of the reshuffled nodes equaled their empirical rank in only $2.3\%$ of the cases on average. Furthermore, $19\%$ of the nodes obtained significantly high or low rankings ($p < 0.1$). These synthetic results demonstrate that our net effects measure is sensitive to the precise neighborhood rewiring performed by the Milisav et al. (2025) algorithm, provided the network size affords a sufficiently large combinatorial state space.\\

\textbf{Null model 2.} To provide a more flexible baseline that relaxes the in and out degree sequences constraint, we implemented a sign-reshuffling null model. This model strictly preserves the empirical unweighted topology (who interacts with whom) and the global ratio of positive to negative edges, but randomly permutes the signs across the existing interactions for each node.
Under this null model, the resulting hierarchical ranks are evenly distributed, see Fig. S3. This finding confirms that a monk’s rank within the monastery hierarchy is not driven merely by their structural connectivity (their underlying unweighted degree), but rather by the specific, strategic arrangement of their positive and negative social ties.

\begin{figure}
    \centering
    \includegraphics[width=\linewidth]{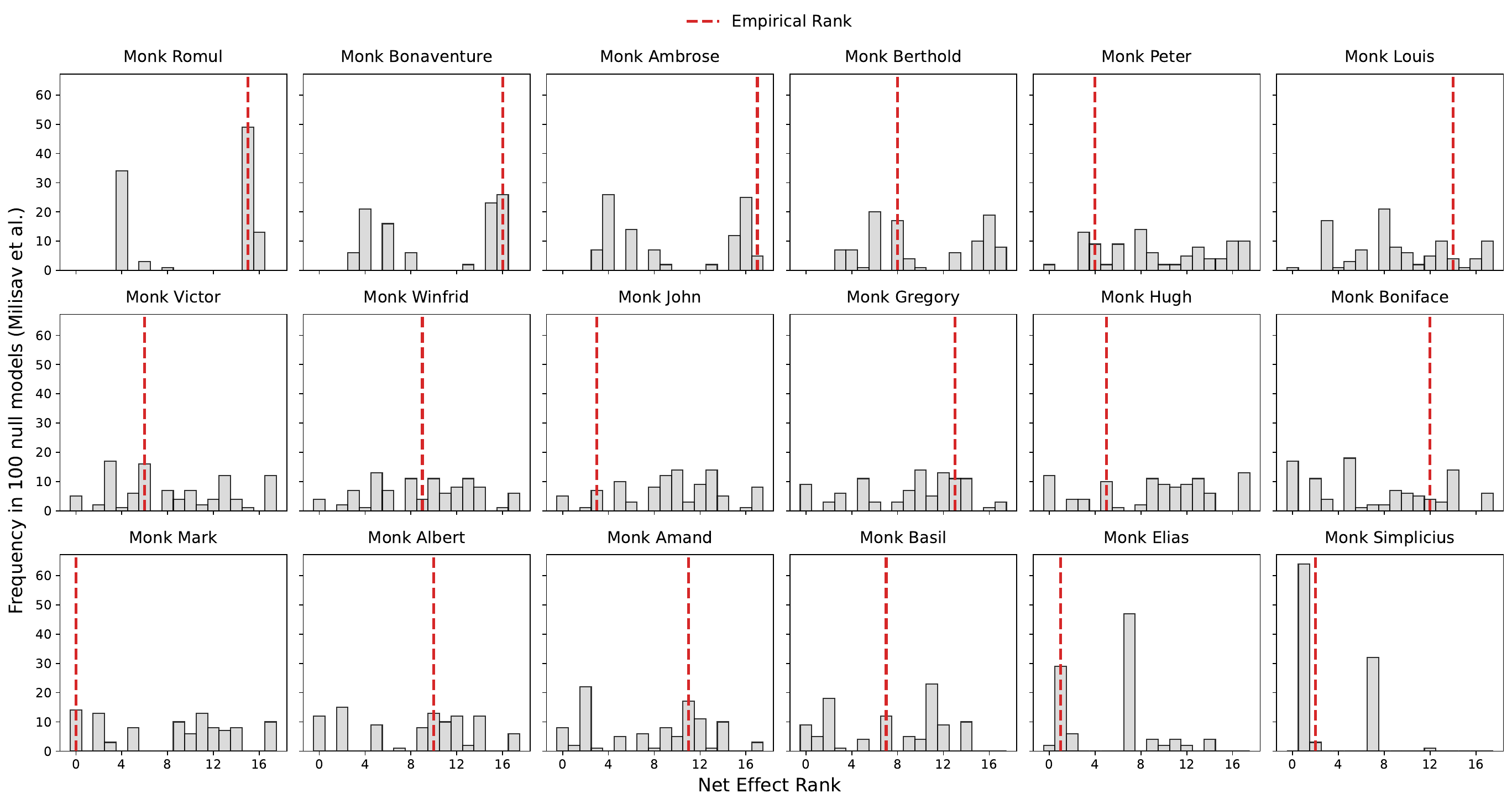}
    \caption{Frequency in 100 randomizations according to Milisav et al. (2025), null model 1, of the ranks of each monk according to the measure of row-scaled outgoing net effects $\mathrm{Net} (\alpha \mathbf{D}^{-1} \! \mathbf{A},\mathrm{all} \leftarrow *)$ for $\alpha = 0.99 < 1$. The rank of the original empirical network is highlighted in red. }
    \label{fig:placeholder}
\end{figure}

\begin{figure}
    \centering
    \includegraphics[width=\linewidth]{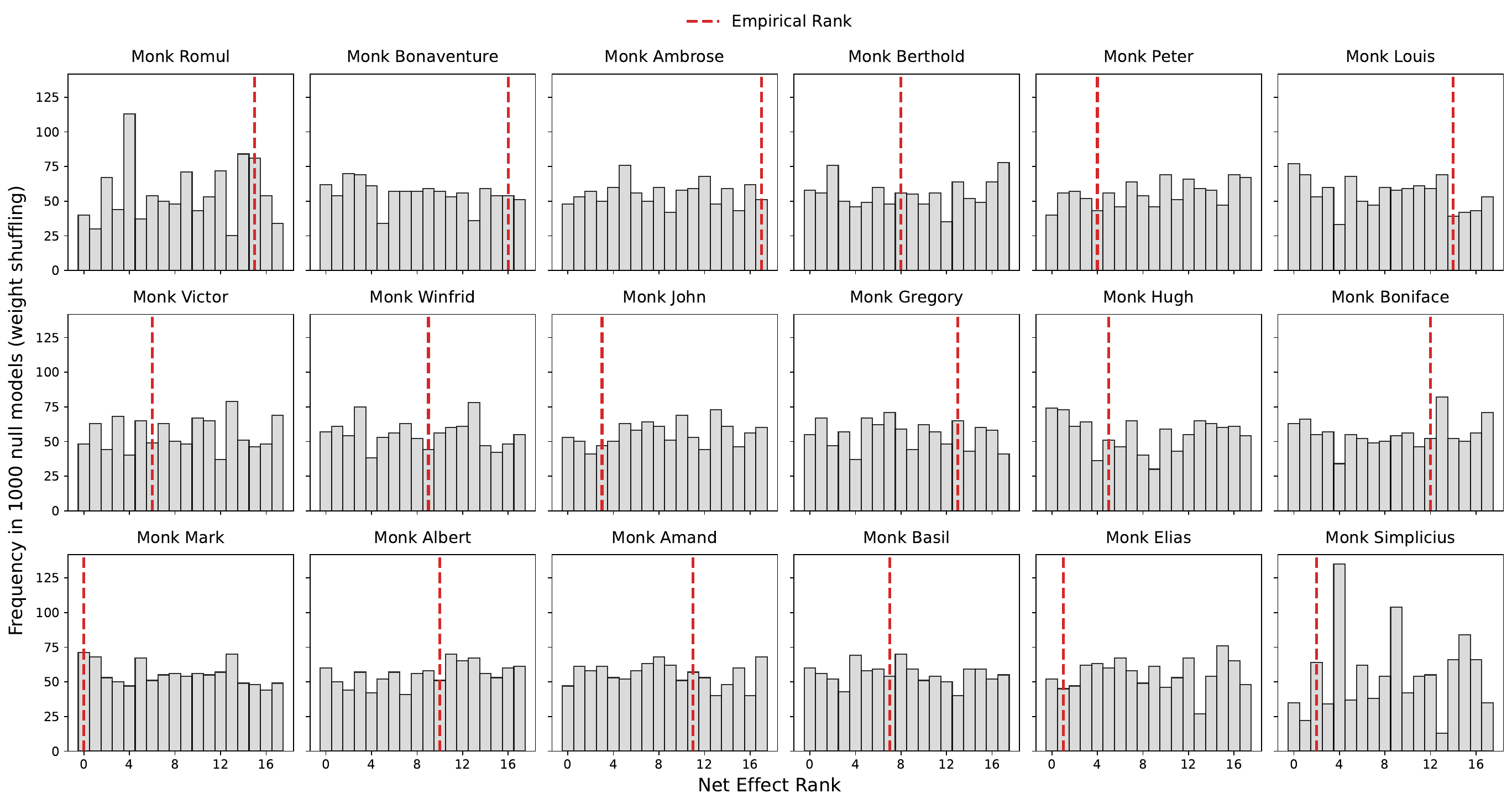}
    \caption{Frequency in 1000 randomizations implementing null model 2 of the ranks of each monk according to the measure of row-scaled outgoing net effects $\mathrm{Net} (\alpha \mathbf{D}^{-1} \! \mathbf{A},\mathrm{all} \leftarrow *)$ for $\alpha = 0.99 < 1$. The rank of the original empirical network is highlighted in red.}
    \label{fig:placeholder}
\end{figure}

\bibliography{references}